\documentclass[aps,twocolumn,prd,nofootinbib,showpacs]{revtex4}
\usepackage[pdftex,bookmarks]{hyperref}

\usepackage{amsfonts}
\usepackage{amsmath,amssymb}
\usepackage[T1]{fontenc}
\usepackage{mathrsfs}
\usepackage{epsfig,epstopdf}

\usepackage[absolute]{textpos}

\usepackage{graphics,wrapfig,times}
\usepackage{graphicx}

\usepackage{color}
\definecolor{D}{rgb}{0.00,0.17,0.48}
\definecolor{M}{rgb}{0.00,0.02,0.83}
\definecolor{L}{rgb}{0.58,0.79,1.00}
\definecolor{R}{rgb}{1,0,0}
\definecolor{B}{rgb}{0.00,0.00,0.00}
\definecolor{P}{rgb}{0.00,0.30,0.60}
\definecolor{W}{rgb}{1,1,1}

\TPGrid[40mm,40mm]{15}{25}  

 \makeatletter
 \newcommand{\pback}[1]{{
   \let\@rrow=\leftarrowfill
   \mathchoice{\AIN@stemPullBack{#1}{\@rrow}}{\AIN@stemPullBack{#1}{\@rrow}}
     {\AIN@indxPullBack{#1}{\@rrow}}{\AIN@indxPullBack{#1}{\@rrow}}}
   \vphantom{#1}}

 \newcommand{\AIN@stemPullBack}[2]{
   \vtop{\mathsurround=0pt
   \ialign{##\crcr$\textstyle{#1}\strut$\crcr
     \noalign{\kern-0.4ex\nointerlineskip}{\tiny#2}\crcr}}}

 \newcommand{\AIN@indxPullBack}[2]{
   \vtop{\mathsurround=0pt
   \ialign{##\crcr\hfil$\scriptstyle{#1}$\hfil\crcr
     \noalign{\kern+0.4ex\nointerlineskip}{\tiny#2}\crcr}}}

\def\bar{\overline}
\def\be{\begin{equation}}
\def\ee{\end{equation}}
\def\bea{\begin{eqnarray}}
\def\eea{\end{eqnarray}}
\def\ba{\begin{array}}
\def\ea{\end{array}}

\def\={\hateq}
\def\puto#1{\rlap{\raise.5ex\hbox{\char'27}}{#1}}

\newcommand{\nn}{\nonumber}
\newcommand{\half}{\frac{1}{2}}

\def\a{\alpha}
\def\b{\beta}

 \def\tt2{{\tilde{2}}}

\def\.{\cdot}

\def\Re{{\rm Re}}
\def\Im{{\rm Im}}
\def\l{\ell}

\def\be{\begin{equation}}
\def\ee{\end{equation}}
\def\bea{\begin{eqnarray}}
\def\eea{\end{eqnarray}}
\def\ba{\begin{array}}
\def\ea{\end{array}}

\newcommand{\eqhat}{\mathrel{\widehat\mathalpha{=}}}
\def\={\eqhat}

\begin{document}

\title{How can one observe gravitational angular momentum radiation from a dynamical source near null infinity?
} 

\author{Chih-Hung Wang}\email{chwang@phy.ncu.edu.tw}
\affiliation{Leichung Waldorf high school, Taichung, 406, Taiwan
}
\author{Yu-Huei Wu}\email{yhwu@phy.ncu.edu.tw}
\affiliation{Center for Mathematics and Theoretical Physics, National Central University, Chungli, 320, Taiwan.}

\begin{abstract}

To answer a question of how can one observed angular momentum radiation near null infinity, one can first transform the dynamical twisting vacuum solution and make it satisfy Bondi coordinate conditions in the asymptotical region of the null infinity. We then obtain the Bondi-Sachs news function and also find the relationship of how does the angular momentum contribute to the news functions from the exact solution sense. By using the Komar's integral of angular momentum, the gravitational angular momentum flux of the dynamical twisting space-time can be obtained. All of our results can be compared with the Kerr solution, Robbinson-Trautmann or Schwarzschild solution. This study can provide a theoretical basis to understand the correlations of gravitational radiations near a rotating dynamical horizon and null infinity.

\end{abstract}

\pacs{04.20.Ha, 04.20.Jb, 04.20.Gz,97.60.Lf}

 \maketitle


\textbf{Introduction and motivation.--} A theoretical framework of studying gravitational outgoing radiations and mass loss at null infinity was originally established by Bondi {\it et al} \cite{Bondi-Sachs} and further developed by Newman and Unit \cite{NU}. In asymptotically  flat empty space-times with outgoing radiation condition and certain coordinate conditions satisfying, the framework provides a systematical analysis to study gravitational energy flux and also non-linear effects of gravitational radiations for general asymptotically flat solutions of Einstein field equations. Besides the study of asymptotical behaviors near the null infinity, the authors apply asymptotical expansions to investigate gravitational radiations on the neighborhood of another space-time boundary, horizon \cite{Wu2007} \cite{Wu-Wang}, however, the authors do not satisfy with the setting of the slow rotation approximation.   The gravitational energy flux and angular-momentum flux across a slowly rotating dynamical horizons (DHs) were obtained in  \cite{Wu2007} \cite{Wu-Wang}.

The merger of two black holes (BHs) is an important source to generate gravitational waves, and it is normally accompanied by the recoil velocity and spin flip \cite{Merritt-Ekers-02} \cite{Baker-06}. In the study of recoil velocity and spin precession during the binary BHs merger, we need to understand not only gravitational radiations near DHs but also gravitational waves propagating to null infinity. It is physically important to established the correlations of geometric quantities between the DHs and null infinity. In \cite{Macedo-Saa-08}, they used Robinson-Trautman (RT) space-time \cite{RT62}, which is an exact solution of Einstein field equations containing spherical GWs, to study gravitational wave recoils. Moreover, Rezzolla, Macedo and Jaramillo \cite{Rezzolla-et-al-10} study the antikick of head-on collision of two nonspinning BHs, which has been observed from numerical-relativity calculations, in RT space-times.  Although RT space-time is an algebraic special solution, i.e. shear-free, it is a dynamical solution and has GWs propagating to null infinity. Bondi coordinate is a physical coordinate which allows us to study the gravitational radiation. Therefore, we must transform it to the Bondi coordinate and the news function appears. The asymptotical behavior of RT space-time in the Bondi coordinate has been first investigated by Foster and Newman \cite{Foster-Newman67}, and they only considered small deviation from spherical symmetry. In \cite{Gonna-Kramer-98}, Gonna and Krammer study the pure and gravitational radiations of RT space-time and obtain Bondi-Sachs news function,  i.e. the gravitational free data.

It is interesting to generalize their works to include the spins in the merger of binary BHs. So, how can one define angular momentum of a dynamical source? Since it is not proper to use RT space-time to study binary BHs merger with spins, the dynamical twisting vacuum solution \cite{SKMHH-03}, i.e. hypersurface non-orthogonal,  may provide a good candidate to study spinning BHs since the Kerr black hole is a stationary vacuum solution of the twisting space-times.

In this paper, we perform a coordinate transformation on twisting vacuum solutions to the Bondi coordinates and apply Newman-Unti (NU) asymptotical expansion \cite{NU} to obtain news function of twisting space-times. One can clearly see how does the angular momentum contribute to the gravitational radiation (the news function). We use the twisting solution of P. 437 Sec. 29 \cite{SKMHH-03} in the complex coordinate $(u, r, \zeta, \bar\zeta)$ and later transform it to the Bondi complex coordinates $(U, R, \zeta', \bar\zeta')$. By writing a twisting vacuum solution in the Bondi coordinate, we then obtain the NU mass and also angular momentum. After obtaining the NU mass, one could combine the news function to obtain the Bondi mass formula. From this work, we may know how the angular parameter contributes to the Bondi mass. We also observe a formula that related with angular momentum and the news function. It is easy to show that when angular parameter vanishes, the solution will return to Robinson-Trautman (RT) spacetime in the Bondi coordinates. The results will be used to study the merger of two spinning BHs and also the influences of the gravitational angular-momentum flux on the spin flip \cite{Wang-Wu 2013b} in the future.

Unfortunately, no explicit expression for the angular momentum in terms of the Kerr parameters $m$ and $a$ is given from the spinor construction of angular momentum, e.g., Bergqvist and Ludvigsen, Bramson's superpotential, Ludvigsen-Vickers angular momentum (See \cite{Szabados-04} for the detail). Thus our angular momentum is calculated by using Komar integral \cite{Komar59}. We show that our calculation will return to Kerr solution and yield the $m a$. Our convention in this paper is $(+---)$. Note that we do not require axial symmetric.


\textbf{Twisting vacuum solutions in the Bondi coordinates.--} Here we use the Newman-Unti affine parameter rather than Bondi luminosity distance. So that our results can be compared with the results of Newman-Unti\cite{NU} or Foster-Newman \cite{Foster-Newman67}. The covariant tetrad of twisting spacetime in $(u, r, \zeta,\bar\zeta)$ coordinate \cite{SKMHH-03} is
\bea \l_a &=& (1, 0, L, \bar L),\; n_a = (H, 1, A, \bar A),\\
  m_a &=& (0,0, 0,-B),\;  \bar m_a =(0,0, -\bar B,0), \eea
where $ A := W+ LH$ and $B :=\frac{1}{\eta P}$ and $A, B, \eta$ are complex and are functions of $(u,r,\zeta,\bar\zeta)$ and $L, W$ are functions of $(u,\zeta,\bar\zeta)$. We then have the NP coefficients and since the spacetime is algebraic special for $\l$, we have $\kappa=\sigma=\lambda=0$ and
%
 \bea H &:=& \frac{K}{2} -r(\ln P)_{,u} - \frac{mr}{r^2+\varpi^2}, \nn\\
      K &:=& 2 P^2\Re [\partial(\bar\partial \ln P -\bar L_{,u})],\;  2i\varpi := P^2 (\bar\partial L-\partial \bar L),\nn \\
      W &=& \frac{L_{,u}}{\eta} +i \partial \varpi,\;  \partial \equiv \partial_\zeta - L\partial_u,  \nn\\
      \eta^{-1} &:=& - (r + i \varpi),\;\;      (B\bar B)^{-1} = \eta\bar\eta P^2 = \frac{P^2}{r^2} + O(\frac{1}{r^3}),\nn\eea
 where we use $r_0=0, M=0$ and $\varpi, P$ is real in \cite{SKMHH-03}.

 The contravariant tetrad in $(u,r,\zeta,\bar\zeta)$ coordinate is
\bea \l^a &=& (0,1, 0,0),\;  n^a = (1, -H, 0,0),\\
      m^a &=& (-\frac{L}{\bar B}, \frac{ W}{\bar B}, \frac{1}{\bar B},0),\;    \bar m^a = (-\frac{\bar L}{B}, \frac{\bar W}{B}, 0, \frac{1}{B}).\eea

 We need to perform a coordinate transformation $\tilde{g}^{ab} =\frac{\partial X^a}{\partial x^i} \frac{\partial X^b}{\partial x^j} g^{ij}$  to transform coordinate $(u, r, \zeta,\bar\zeta)$ to Bondi coordinate $(U, R, \zeta', \bar\zeta')$.
 \bea U &=& U_0 +\frac{U_1}{r} + O(\frac{1}{r^2}),\\
    R &=& R_{-1}r + R_0 + \frac{R_1}{r}+ O(\frac{1}{r^2}),\\
    \zeta' &=& \zeta_0 +\frac{\zeta_1}{r} + O(\frac{1}{r^2}),\;
    \bar\zeta' = \bar\zeta_0 +\frac{\bar\zeta_1}{r} + O(\frac{1}{r^2}).\eea

 The metric of the twisting spacetime is
 \bea  g^{00} &=& g^{uu} = -2\frac{\bar L L}{\bar B B} =\frac{g^{00}_0}{r^2} + O (\frac{1}{r^3}), \nn \\
       g^{01} &=& 1 + \frac{2 H L \bar L - L\bar A -\bar L A}{ B \bar B} = 1+ \frac{g^{01} _0}{r}+ O(\frac{1}{r^2}), \nn\\
       g^{02} &=& \frac{\bar L}{\bar B B} = \frac{g^{02}_0}{r^2} +O(\frac{1}{r^3}),\;
       g^{03} = \frac{L}{\bar B B} = \frac{g^{03}_0}{r^2} +O(\frac{1}{r^3}),\nn\\
       g^{11} &=& -2 H -2 \frac{A\bar A - AH \bar L -\bar A H L + H^2 L\bar L}{B\bar B}\nn\\
        &=& -\tilde{K}+ \frac{2 \tilde{m}}{r} + r\partial_u \ln P+ O(\frac{1}{r^2}),\nn\\
       g^{12} &=&\frac{\bar W}{B \bar B} =\frac{g^{12}_0}{r} +O(\frac{1}{r^2}),\nn\\
       g^{13} &=&\frac{W}{B \bar B} =\frac{g^{13}_0}{r} +O(\frac{1}{r^2}),\nn\\
       g^{23} &=& g^{\zeta\bar\zeta} = -\frac{1}{B\bar B}= \frac{g^{23}_0}{r^2} +O(\frac{1}{r^3}).\nn\eea
 where
 \bea g^{01}_0 &=& -P^2 (L\bar L)_{,0},\;\; g^{02}_0 = P^2 \bar L,\;\; g^{03}_0 = P^2 L,\nn\\
      g^{12}_0 &=& -P^2 \bar L_{,0},\;\; g^{13}_0 = -P^2 L_{,0},\nn\\
      g^{00}_0 &=& -2 \bar L L P^2,\;\; g^{23}_0 = - P^2,\nn\\
      \tilde{K} &=& K -2 P^2 |L_{,0}|^2,\nn\\
      \tilde{m} &=& m + P^2 [-2 \Sigma L_{,0} \bar L_{,0} + L_{,0} \bar\partial \Sigma +\bar L_{,0} \partial \Sigma].\nn\eea
and the asymptotic values of $H, B, W$ are
\bea H&=& \half K -\frac{m}{r} - r(\ln P)_{,u} +O(\frac{1}{r^2}),\\
     B&=&-\frac{r}{P}  +O(1),\\
     W&=& -L_{,0} r + i (- \Sigma L_{,0} + \partial \Sigma),\eea
where $L= L(u, \zeta,\bar\zeta)$, $P =P(u,\zeta,\bar\zeta)$,and $\Sigma =\Sigma(u,\zeta,\bar\zeta)$.

Bondi coordinate conditions are
\bea \tilde{g}^{00} = \tilde{g}^{02} = \tilde{g}^{03} = 0,\;\;
\tilde{g}^{01} = -1 ,
\eea
which we choose affine parameter here. 

\textbf{Transform Newman-Unti real coordinate $(U, R, x^2, x^3)$ to Bondi complex coordinate $(U,R,\zeta',\bar\zeta')$.--}
Here we use a complex coordinate $(\zeta,\bar\zeta)$ to make our whole calculation simpler. We need to use a coordinate transformation to transfer Newman-Unti real coordinate $(x^2,x^3)$ \footnote{ Newman-Unti originally use $(x^3,x^4)$, here our coordinate runs from $(x^0, x^1, x^2, x^3)$.} into the complex coordinate $(y^2,y^3)=(\zeta',\bar\zeta')$ in order to compare with Newman-Unti \cite{NU}. We have
\bea y^2=\zeta'=\half( x^2+i x^3),\;\;\; y^3 = \bar\zeta'= \half (x^2-i x^3).\eea
 %
From $\tilde{g^{\a\b}} =\frac{\partial y^\a}{\partial x^m} \frac{\partial y^\b}{ \partial x^n}  g^{mn}_{NU}$, we get
\bea \tilde{g^{11}} &=& \frac{\partial y^1}{\partial x^m} \frac{\partial y^1}{ \partial x^n} g^{mn}_{NU}= g^{11}_{NU},\\
  \tilde{g^{01}} &=& 1,\;\; \tilde{g^{12}} = \half g^{12}_{NU} + i \half g^{13}_{NU},\\
  \tilde{g^{13}} &=& -i\half g^{13}_{NU} + \half g^{12}_{NU},\\
  \tilde{g^{22}} &=& \frac{1}{4} g^{22}_{NU} +\frac{1}{4} g^{33}_{NU} +i\half g^{23}_{NU}\\
   &=& 2 P^2_{NU} \sigma^0 R^{-3}+O(R^{-4}),\\
   \tilde{g^{23}} &=& \frac{1}{4} g^{22}_{NU} -\frac{1}{4} g^{33}_{NU} \\
   &=& - P^2_{NU} R^{-2} -3 \sigma^0 \bar\sigma^0 P^2_{NU} R^{-4}+ O(R^{-5}),\\
   \tilde{g^{33}} &=&  \frac{1}{4} g^{22}_{NU} -\frac{1}{4} g^{33}_{NU} -i\half g^{23}_{NU}\\
   &=& 2 P^2_{NU} \bar\sigma^0 R^{-3}+O(R^{-4}),
      \eea
and we use the results of Newman-Unti \cite{NU}
\bea
g^{11}_{NU} &=& -2 P_{NU}^2 \nabla\bar\nabla \ln P_{NU} - (\Psi^0_2+\bar\Psi^0_2) R^{-1} +O(R^{-2}),\nn\\
g^{22}_{NU} &=& -2 P^2_{NU} R^{-2} + 2 P^2_{NU} (\sigma^0+\bar\sigma^0) R^{-3} +O(R^{-4}),\nn\\
g^{23}_{NU} &=& - 2i P^2_{NU} (\sigma^0-\bar\sigma^0) R^{-3} + O(R^{-4}),\nn\\
g^{33}_{NU} &=& -2 P^2_{NU} R^{-2} - 2 P^2_{NU} (\sigma^0+\bar\sigma^0) R^{-3}+ O(R^{-4}),\nn \eea
where $ P_{NU} =P_{NU}(x^2,x^3)$ and later we will prove $f=P_{NU}$. Note that we first introduce $f$ from $O(r)$ of $\tilde{g}^{11}$. The definition of Newman-Unti mass integral is $M_{NU}:=\oint m_{NU} dS$ where the Newman-Unti mass is defined as $m_{NU}:= \Re\Psi_2^0$.

\textbf{Results from coordinate transformation.--} From $\tilde{g}^{00}$ the $O(r^{-2})$ term vanishes, we get
     \bea U_1 = -f P U_{0,2} U_{0,3} - \frac{P^3 L\bar L}{f} +P^2\bar L U_{0,2} + P^2 L U_{0,3}.\nn\eea
From $\tilde{g}^{01} =1 $ and $O(1)$ vanishes, we get
     \bea R_{-1} =U_{0,0}^{-1}\eea
     From $O(r^{-1})$ term vanishes, we obtain an identity:

From $\tilde{g}^{02} = O(r^{-3})$ and $\tilde{g}^{03} = O(r^{-3})$ , we get
     \bea \zeta_1 &=& P(P\bar L_0 - f U_{0,3}),\;\;  \bar\zeta_1 = P(P L_0 - f U_{0,2}).\eea
From $\tilde{g}^{R\zeta'} = \tilde{g}^{12} =\tilde{g}^{13} = O(r^{-1})$, we get
      \bea \zeta_{0,0} = 0,\; \bar\zeta_{0,0}=0,\;
            \zeta_{0,2}=1,\; \bar\zeta_{0,3}=1, \eea
and $\zeta_0=\zeta$, $\bar\zeta_0=\bar\zeta$.

From $O(r)$ of $\tilde{g}^{RR} =\tilde{g}^{11}$, we obtain a differential equation $(\ln R_{-1} +\ln P +K)_{,0} =0$ and thus get
 \bea R_{-1} = f P^{-1}\eea
 where $f$ is a function that depends on $(\zeta,\bar\zeta)$. Thus, $U_{0,0} =\frac{P}{f}$.

 From $O(1)$, we obtain $R_{0,0}$.
  After integration, we can further obtain
  \bea R_0 =f^2 U_{0,23} +\bar L (f_{,2} P - f P_{,2}) + L(f_{,3} P - f P_{,3}) \nn\\
  -\half f P (\partial_2 \bar L + \partial_3 L) + f P_{,0} L \bar L - f P (L\bar L)_{,0}.\eea

 From $O(r^{-1})$, we obtain the \textit{Newman-Unti mass} $m_{NU} = \half (\Psi^0_2+\bar\Psi^0_2)=\Re \Psi^0_2$ after integration,
 \bea m_{NU} &=& -2P L\bar L R_{-1,0} R_{0,0} + \frac{R_{1,0} R_{-1} - R_{-1,0} R_{1}}{P} \nn \\
 &-& P(L\bar L)_{,0} R_{0,0} R_{-1} \nn\\
 &+& [P \bar L (R_{-1,0} R_{0,2} + R_{0,0} R_{-1,2})  + C.C]\nn\\
   &-&2 \frac{P_{,0} R_{-1} R_1}{P^2} + \frac{R_{-1}^2}{P} \tilde{m} - [R_{-1} R_{0,2}\bar L_{,0} P + C. C.]\nn\\
   &-& (R_{-1,2} R_{0,3} + R_{0,2} R_{-1,3}) P. \label{Mnu}\eea
 which we need $R_0,R_{-1}, R_1$.
 From $\tilde{g}^{23} =\tilde{g}^{\zeta'\bar\zeta'}$,
 we get \bea f = P_{NU}\eea from $O(r^{-2})$.
 From $O(r^{-3})$, we obtain $R_0$ which yield the same result with the one from $\tilde{g}^{RR}$.
 From $\tilde{g}^{22} =\tilde{g}^{\zeta'\zeta'}$ and $O(r^{-3})$, we obtain \textit{the shear term}.
 \bea \sigma^0 &=& \frac{f}{P} (P^2 \bar L)_{,3} + \frac{P f \bar L_{,0}^2}{2} + 2 \frac{f^2}{P} P_{,0} U_{0,3} \bar L \\
  &+& (f^2 U_{0,3})_{,3} - \frac{f^3}{P^2} (U_{0,3})^2.\eea
One can calculate the Bondi news function by applying $\partial_u$
\bea \dot{\bar{\sigma^0}} &=&\frac{\partial}{\partial u}"U_{0,2} \;\;\textrm{terms} "+ \frac{\partial}{\partial u} "L\;\; \textrm{terms}"\nn\\
 &=& f(\frac{(P^2 L)_{,2}}{P})_{,0} + \frac{f}{2} (P L_{,0}^2)_{,0} + 2 f^2 (\frac{P_{,0} U_{0,2} L}{P})_{,0}\nn\\
 &+&[(f^2)_{,2} U_{0,2} + f^2 U_{0,22}]_{,0} - f^3[\frac{P_{,0}(U_{0,2})^2}{P^2}]_{,0},  \label{shear}\eea
 and \textit{the Bondi News function} is defined as $\frac{\partial}{\partial U}{\bar{\sigma^0}}= \frac{P}{f} \frac{\partial}{\partial u}{\bar{\sigma^0}} $, thus one can work out \textit{the Bondi mass} $\Psi_2^0+\sigma^0 \frac{\partial}{\partial U}\bar\sigma^0$ from $m_{NU}$ by using (\ref{Mnu}) and (\ref{shear}). The mass loss comes from the news function $\frac{\partial}{\partial U}{\bar{\sigma^0}}$. When $L=0$, the results should return to the \cite{Gonna-Kramer-98} written by using spherical coordinate. Also, $\Psi^0_4 =-\frac{\partial^2}{\partial U^2}\bar\sigma^0$ can be calculated.

\textbf{Komar's angular momentum.--} From $x^2 =\theta, x^3 =\phi$ and $\zeta=\sqrt{2} e^{i \phi} \cot \frac{\theta}{2}$ \cite{Penrose}, we have
$\frac{\partial}{\partial \zeta} = Q^2 \frac{\partial}{\partial \theta} + Q^3 \frac{\partial}{\partial \phi}$
where $Q^2 = -\frac{1}{\sqrt{2}} e^{-i\phi} \sin^2\frac{\theta}{2}$ and $Q^3 = -i\frac{1}{2 \sqrt{2}} e^{-i\phi} \tan\frac{\theta}{2}$.
The rotation vector (asymptotically Killing vector) $\partial/\partial \phi$ is
\bea \frac{\partial}{\partial \phi}=\phi^a =- i\frac{3}{4} (\zeta\frac{\partial}{\partial \zeta} -\bar\zeta\frac{\partial}{\partial \bar\zeta})\eea
where $\frac{\partial}{\partial \zeta} = \bar B \tilde{m}^a + L \tilde{n}^a + A \tilde{\l}^a$. Note that "$\tilde{}$" represents the null rotation that make $m,\bar m$ tangent for two sphere on null infinity.
We obtain the Komar angular momentum for the twisting space time \cite{Komar59} is
\bea J_K &=& \frac{1}{2\pi} \oint \nabla^a \tilde{\phi}^b d \tilde{S_{ab}}\\
         &=& \frac{1}{2\pi} \oint i \frac{-3}{2}(\zeta L-\bar\zeta\bar L )\Re \tilde{\gamma} d S \\
         & \approx& \frac{1}{2\pi} \int i\frac{3}{4} (\zeta L-\bar\zeta\bar L ) m_{NU} d \theta d \phi
         \eea
where $d S_{ab}= l_{[a} n_{b]} d S$, "$\approx$" represents approximate on null infinity,  $d S \approx r^2 \sin\theta d \theta \phi$, $m_{NU}:= \Re \tilde{\Psi^0_2}$ and we use the results of Newman-Unti \cite{NU} $\gamma=-\frac{\Psi^0_2}{2 r^2}+O(r^{-3})$. This angular momentum will yield $m a$ for Kerr solution.

\textbf{Conclusions.--} We have build up the relationship of how the angular momentum contribute to the gravitational radiation (the news function) from the exact solution (the twisting space-time).   Dynamical twisting space-time is a solution that allows the freedom of gravitational radiation from the exact solution sense and it characterizes spin.  
Therefore, we must transform it to the Bondi coordinate and the news function appears. Though there is no satisfactory way to have an explicit expression for the angular momentum in terms of the Kerr parameters from spinor construction. However, we use Komar integral for the twisting space-time and get a general expression that is related with NP $\Re \gamma$. $\Re\gamma$ is something like the surface gravity of horizon $\Re \epsilon$ for $\l$. Further from the results of Newman-Unti, it can be rewritten as $\Re\Psi_2$, i.e., the Newman-Unti mass term. Thus it would be very easy to check that our results should go back to the Kerr solution. We hope that these results will help us to understand the dynamics of the merger of  two spinning BHs and see the influences of the gravitational angular momentum flux on the spin flip problem  which we have write down the initial state and the final state of the merge \cite{Wang-Wu 2013b}. Also, we have worked out Komar angular momentum for Kerr black hole in \cite{Wu2007} and for null infinity in this paper. It would be interesting to build up a further correlation between horizon and null infinity.


\textbf{Acknowledgment.--} The key ideas of this paper were carried out by YH Wu during the visit of Albert Einstein Institute (AEI), Golm, July 2013. 
 YH Wu and CH Wang would like to express deep gratitude to Dr. Jose-Luis Jaramillo for academic discussion and  AEI for hospitality and travel support. CH Wang would like to thank GR 20 conference center for the travel support. YH Wu would like to thank Prof. Yun-Kau Lau to first draw her attention to this problem.

\textbf{Appendix: Kerr solution and its angular momentum.--} To calculate Komar's angular momentum, one needs to make $m, \bar m$ tangent on two sphere at null infinity for Kerr solution. Firstly, we can write down the Kerr solution in Bondi coordinate. From the coordinate transformation $k^{a'} = \frac{\partial x^{a'}}{\partial x^a} k^a$, we have
$U_{0,0} = \frac{P_{Kerr}}{f} =1$, then
$U_0 = u $, $U_1 =-\frac{a \zeta\bar\zeta}{P_{Kerr}^2}$, $R_{-1} = \frac{f}{P_{Kerr}} =1$, $R_0 =0$, $\zeta_0 =\zeta $, $\zeta_1 = i a \zeta$.
The leading order of the contravariant tetrad in Bondi coordinate $(U, R,\zeta',\bar\zeta')$ is
\bea \l^a &=&( 0,1, 0,0 ),\;
      n^a = (1, -H_{Kerr}, 0, 0) \nn\\
      m^a &=& (\frac{P_{Kerr}L_{Kerr}}{r}, \frac{P_{Kerr} L_{Kerr}}{ r}, -\frac{P_{Kerr}}{r},0),\nn
\eea
where we ignore the higher order terms $O(r^{-2})$. For the Kerr solution, we use
\bea P_{Kerr} &=& 1+ \half\zeta\bar\zeta= 1+\cot(\frac{\theta}{2})^2,\; K=1,\nn \\
L_{Kerr} &=&- i \frac{a\bar\zeta}{P^2_{Kerr}}=-i\frac{a \sin^2\theta}{2}=-W_{Kerr},\nn\\
\varpi_{Kerr}&=&-a\frac{2-\zeta\bar\zeta}{2+\zeta\bar\zeta},\; B_{Kerr}=-\frac{1}{\eta P_{Kerr}}\nn\\
H _{Kerr}&=& \half -\frac{m r}{r^2 +\varpi_{Kerr}^2}.\nn \eea
Secondly, we null rotate it to make $m, \bar m$ tangent on the two sphere at null infinity. We perform type II null rotation $m'= m + b n$,$\l'=\l +\bar b m +b \bar m+b\bar b n$, and $n'=n$ where
$ b= -\frac{P_{Kerr} L_{Kerr}}{r},$
and it satisfy the gauge conditions $\pi=-\bar\tau, \Im \rho=0, \Re \epsilon =0.$ Thus, after null rotation, the Komar angular momentum for Kerr solution is $m a$ and we also check this point with GRtensor Maple.


\begin{thebibliography}{10}



\bibitem{Bondi-Sachs} Bondi H  van der Burg M G  J and  Metzner  A W K,  Proc  Roy Soc (London) \textbf{A269} 21 (1962).
Sachs, R. K, 
 Proc. Roy. Soc (London), A264, 309-338 (1961).


\bibitem{Foster-Newman67} J. Foster and E. T. Newman, J. Math. Phys. \textbf{8},189 (1967).

\bibitem{Komar59} Komar, A., 
 Phys. Rev., 113, 934-936,
(1959).


\bibitem{NU}  Newman , E. T. and  Unit, T. W. J., J. Math. Phys. \textbf{3}, 891-901 (1962).

\bibitem{Baker-06} Baker, J. et al. , 
 Phys. Rev. Lett. 96 111102 2006

\bibitem{Merritt-Ekers-02} D. Merritt and R. Ekers, 
 Science 297, 1310 (2002);



\bibitem{Szabados-04} Szabados L B 2004,
Living Rev. Rel. 7. 4. 



\bibitem{Penrose} Penrose, R., and Rindler, W., Spinors and space-time, Vol. 1 and Vol.2
    (Cambridge University Press, Cambridge; New York, 1984 and 1986).




\bibitem{Macedo-Saa-08}    R. P. Macedo and A. Saa, Phys. Rev. D 78, 104025 (2008).

\bibitem{Gonna-Kramer-98} U von der G\"{o}nna and D Kramer  1998 Class. Quantum Grav. 15 215

\bibitem{RT62} I. Robinson and A. Trautman, Proc. R. Soc. A 265, 463 (1962).

\bibitem{Rezzolla-et-al-10} L. Rezzolla, R. P. Macedo and J. L. Jaramillo, Phys. Rev. Lett. 104,221101 (2010).

\bibitem{SKMHH-03}H. Stephani, D. Kramer, M. A. H. MacCallum, C. Hoenselaers, and E. Herlt,
Cambridge University Press (2003).

\bibitem{Wu2007} Yu-Huei Wu, PhD thesis, University of Southampton (2007).

\bibitem{Wu-Wang} Yu-Huei Wu and Chih-Hung Wang,
    Phys. Rev. D 83, 084044 (2011). Yu-Huei Wu and Chih-Hung Wang, 
    Phys. Rev. D 80, 063002 (2009).

\bibitem{Wang-Wu 2013b} Wang and Wu, 
paper in preparation (2013). Yu-Huei Wu, "Understand law of precession during the merger of binary black hole", GR20, Poland.

\end{thebibliography}
\end{document}